\documentclass[aps,pra,superscriptaddress,twocolumn,showpacs]{revtex4}
\usepackage{graphicx}
\usepackage{amssymb}
\usepackage{amsmath}
\usepackage{bm}

\usepackage{tikz}
\usetikzlibrary{shapes, positioning}

\usepackage{color}

\begin{document}
\setcounter{page}{0}

\title[]{Information-theoretic analysis of complex eigenmodes across avoided crossings in open microcavities}
\author{Kyu-Won \surname{Park}}
\email{parkkw7777@gmail.com}
\affiliation{Department of Mathematics and Research Institute for Basic Sciences, Kyung Hee University, Seoul, 02447, Korea}

\author{Soojoon \surname{Lee}}
\affiliation{Department of Mathematics and Research Institute for Basic Sciences, Kyung Hee University, Seoul, 02447, Korea}
\affiliation{School of Computational Sciences, Korea Institute for Advanced Study, Seoul 02455, Korea}

\author{Kabgyun \surname{Jeong}}
\email{kgjeong6@snu.ac.kr}
\affiliation{Research Institute of Mathematics, Seoul National University, Seoul 08826, Korea}
\affiliation{School of Computational Sciences, Korea Institute for Advanced Study, Seoul 02455, Korea}

\date{\today}

\begin{abstract}
Avoided crossings (ACs) are hallmark signatures of mode interaction in quantum and wave systems. Open microcavities whose resonances are naturally described as quasi-normal modes (QNMs) with complex eigenfrequencies offer a convenient platform to observe how openness and loss reorganize modal structure. We introduce a compact \emph{quadrature space} framework that represents a complex QNM by probability weights on its real and imaginary quadratures, and we apply Shannon-type measures to these distributions. This representation separates marginal spreading of each quadrature from inter quadrature correlation and is robust to nodal sets and exterior zero amplitude points. Applying the method to AC regions, we find that delocalization is driven not only by broadening of individual quadratures but also by a pronounced increase in internal correlation near at the AC, revealing an internal reorganization of resonant modes in non-Hermitian settings. The approach is broadly transferable to other open resonator platforms and provides a general information-theoretic diagnostic for openness-driven mode interactions.
\end{abstract}

\maketitle

\section{Introduction}
Avoided crossings are ubiquitous indicators of mode interaction in wave and quantum systems~\cite{vNW29,Zener32,Majorana32}, and they become especially revealing in open settings. In open resonators, the relevant solutions are naturally described as quasi-normal modes (QNMs) with complex eigenfrequencies that satisfy outgoing radiation conditions~\cite{Leung90,Leung94a,Kristensen14}. Dissipation and radiation shift spectra to the complex plane and they promote non-orthogonality~\cite{Rotter09,Moiseyev11}, and as a result avoided crossings appear together with line width exchange and pronounced mode mixing~\cite{Pichugin23,Park18, Dahiya24_EntropicExchange}. These spectral signatures are well documented, yet a compact account of how the eigenmodes themselves reorganize across an interaction remains less developed. Most studies quantify delocalization from the real space intensity distribution, typically relying on Shannon-type measures of $|\psi(x,y)|^{2}$~\cite{Park18,Dahiya24_EntropicExchange,Arranz19,SanchezDehesa22}. These measures are useful for quantifying spatial spreading, yet they leave unresolved the internal organization of QNMs, motivating a framework that treats their quadratures explicitly.

Dielectric microcavities provide a versatile photonic platform in which these questions can be addressed. They support high quality resonances and are inherently non-Hermitian because of openness and material loss~\cite{CaoWiersig15}. Microcavities underpin key functions in photonics, including lasing~\cite{Ge10}, sensing~\cite{Vollmer08}, and on-chip signal processing~\cite{Bogaerts12}, and modal coupling together with dissipation determines thresholds, linewidths, and sensitivity~\cite{Siegman89,Petermann79,Cheng96}. Their tunable geometries enable interacting-mode scenarios that generate avoided crossings~\cite{Ryu09} and, in some regimes, approach exceptional-point behavior~\cite{ElGanainy18,Park20_MaximalShannonEP}. This combination of controllability and observability makes microcavities an effective setting in which spectral indicators of interaction can be connected with changes in the underlying field structure.

However, views based only on intensity cannot reveal how the real and imaginary quadratures of a complex eigenmode co-organize in open systems. The field is intrinsically bi-quadrature, as emphasized in QNM theory~\cite{Leung94b,Lalanne18}, and therefore a descriptor that separates marginal spreading from structural reorganization in a coordinate free manner is desirable. An analysis carried out directly in the quadrature space of the complex eigenmodes themselves, rather than only in physical coordinates, can expose features of the eigenmode that real space intensity cannot provide.

In this work, we develop an information-theoretic framework defined in the quadrature space of complex eigenmodes. We treat the real and the imaginary components as probabilistic variables with weights derived from interior samples, and we compare their quadrature space distributions consistently along a control parameter. When this framework is applied to an open microcavity across an avoided crossing, the analysis shows that the observed delocalization reflects broader marginals and also a structural redistribution that becomes visible in quadrature space. The framework is phase consistent, robust to nodal sets, independent of external references, and readily transferable to other non-Hermitian wave platforms.

The remainder of the paper proceeds as follows. Section~\ref{sec:spectrum_fields} reviews spectral interaction together with field behavior in the open microcavity. Section~\ref{sec:value_space} formulates the \emph{quadrature space} framework and the cross parameter comparison, and  the \emph{quadrature space} discretization and information measures are studies in Section~\ref{sec:quad_space}. Finally, we summarize our result in conclusion in Section~\ref{sec:conclusion}.

\section{Spectral interaction and field decomposition in microcavities}
\label{sec:spectrum_fields}

\subsection{Eigenvalue spectra and their eigenmodes in microcavities}
We study two-dimensional microcavities governed by the scalar Helmholtz equation for TM polarization,
\begin{align}
(\nabla^2 + n^{2}k^2)\,\psi(\mathbf{r}) = 0,
\label{eq:HelmholtzOpen}
\end{align}
where $\psi(\mathbf r)$ is the out of plane electric field ($E_z$), $n(\mathbf r)=n_{\rm in}=3.3$ inside and $n(\mathbf r)=n_{\rm out}=1$ outside. At the dielectric boundary we impose the usual TM interface conditions (continuity of $\psi$ and of $\partial\psi/\partial n$), and in the exterior we enforce the outgoing (Sommerfeld) radiation condition. Consequently, resonant wavenumbers $k$ are complex, $K=K_{\rm r}-i\,K_{\rm i}$ with $K_{\rm i}>0$ encoding radiative loss. We compute resonances and fields using a boundary element method (BEM), i.e., a boundary-integral scheme with outgoing Green’s functions that naturally yields a non-Hermitian effective operator appropriate for open systems~\cite{Wiersig03_BEM}.
In this paper, we parametrize the elliptical cavity by semi-axes
$ a = (1+\varepsilon), b = (1+\varepsilon)^{-1}$,
so that the area \(\pi a b\) remains fixed at \(\pi\) while the shape is controlled by the deformation parameter \(\varepsilon\) (with \(a\) the major semi-axis and \(b\) the minor semi-axis).

\begin{figure}[htbp]
  \centering
  \includegraphics[width=8.0cm]{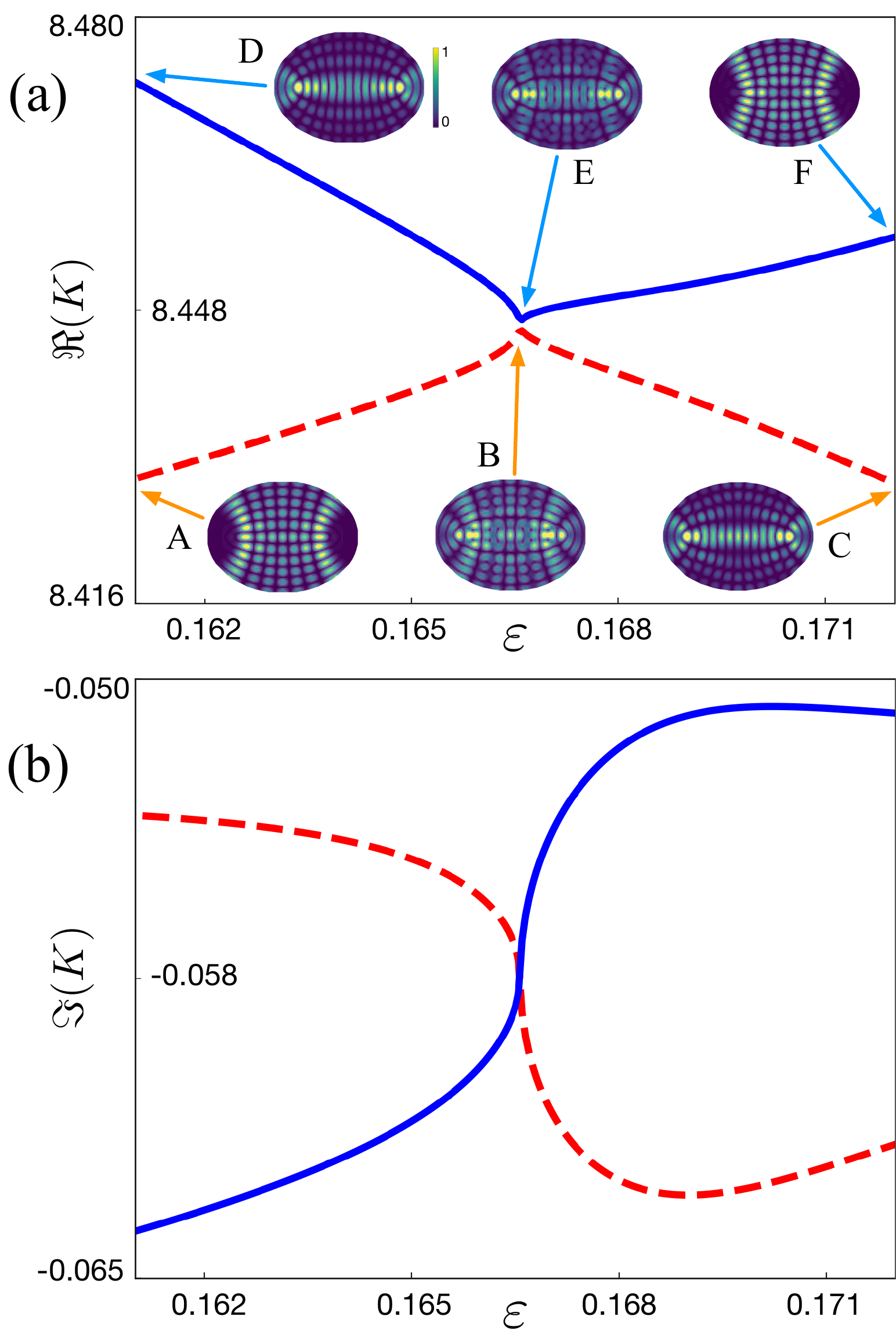}
  \caption{(a) Real parts $\Re(K)$ of two interacting modes plotted against the control parameter $\varepsilon$ (blue solid line: {\bf Mode~2}; red dashed line: {\bf Mode~1}). Insets A--C (bottom row) and D--F (top row) display the spatial intensity distributions at the labeled parameter values. (b) Corresponding imaginary parts $\Im(K)$ over the same parameter range $\varepsilon\in[0.161,0.172]$. The real components exhibit an avoided crossing (AC) near at $\varepsilon\approx0.166$--$0.167$, whereas the imaginary components cross, highlighting the complementary spectral behavior and the pronounced mode reshaping across the interaction.}
  \label{fig:mode_evolution}
\end{figure}
Figure~\ref{fig:mode_evolution} shows the spectral evolution and concurrent spatial reconfiguration of two interacting eigenmodes across the control parameter window $\varepsilon\in[0.161,0.172]$.
Panel (a) plots the real parts $\Re(K)$ of the two eigenmodes (blue solid: {\bf Mode~2}; red dashed: {\bf Mode~1}) and panel (b) shows the corresponding imaginary parts $\Im(K)$. A clear avoided crossing (AC) is observed in the real part trajectories near $\varepsilon\approx0.16655$, while the imaginary parts undergo a crossing in the same region. This complementary spectral response of level repulsion in $\Re(K)$ accompanied by an exchange in $\Im(K)$ is a clear sign of strong mode coupling and the transfer of radiative loss (or linewidth) between the branches.

Insets A--C (bottom row) and D--F (top row) in Fig.~\ref{fig:mode_evolution} display the spatial intensity distributions at the labeled parameter values along the two branches.
Far from the AC each eigenmode maintains a distinct spatial character; approaching the interaction region the patterns hybridize and exhibit comparable intensity on features that were previously mode-selective. At the AC the eigenmodes therefore reach maximal hybridization, consistent with enhanced delocalization of modal weight across the cavity.

It is instructive to place these observations in a minimal coupled-mode framework for open systems. We use the standard $2\times 2$ non-Hermitian effective Hamiltonian
\begin{align}
H_{\rm eff}(\varepsilon)=
\begin{pmatrix}
\Omega_1(\varepsilon) & g\\[3pt]
g & \Omega_2(\varepsilon)
\end{pmatrix},
\label{eq:Heff_simple}
\end{align}
where $\Omega_j(\varepsilon)=\omega_j(\varepsilon)-i\,\gamma_j(\varepsilon)$ with coherent coupling $g\in\mathbb R$ and loss rates $\gamma_j\ge 0$~\cite{Rotter09, Park20_MaximalShannonEP}. Its eigenvalues are given by
\begin{align}
\lambda_\pm(\varepsilon)
=\frac{\Omega_1+\Omega_2}{2}
\ \pm\
\sqrt{\left(\frac{\Omega_1-\Omega_2}{2}\right)^{\!2}+g^2}.
\label{eq:eigs_simple}
\end{align}
For $g\neq 0$, the real parts $\Re(\lambda_\pm)$ generically repel near the detuning $\Omega_1\simeq\Omega_2$ (avoided crossing), while the imaginary parts can cross when the differential loss $|\gamma_1-\gamma_2|$ is not too large compared to $|g|$, reproducing the pattern in Fig.~\ref{fig:mode_evolution}.

While Fig.~\ref{fig:mode_evolution} establishes these spectral features and their impact on intensity patterns A--F, the underlying field rearrangement is clarified by decomposing the complex eigenfields into their \emph{quadrature space} components.

\subsection{Field decomposition in quadrature space}
\begin{figure*}[htbp]
  \centering
  \includegraphics[width=17.0cm]{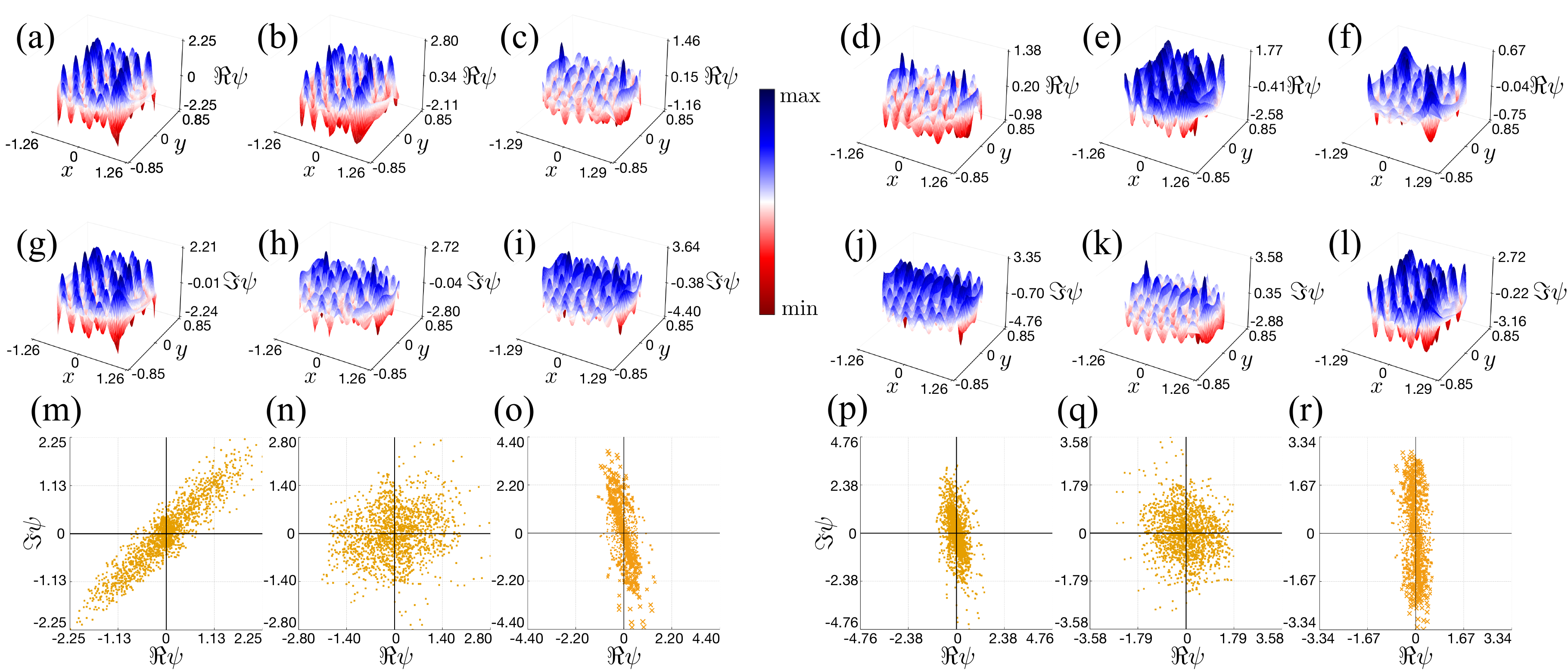}
  \caption{(a--c) Real parts $\Re[\psi(x,y)]$ corresponding to intensity panels A--C of {\bf Mode~1} in Fig.~\ref{fig:mode_evolution}.
  (d--f) Real parts for {\bf Mode~2} corresponding to panels D--F.
  (g--i) Imaginary parts $\Im[\psi(x,y)]$ for {\bf Mode~1} at A--C, and (j--l) Imaginary parts for {\bf Mode~2} at D--F.
  (m--r) Complex-plane representations obtained by plotting weighted scatter points $(\Re\psi,\Im\psi)$ with weights $|\psi|^2$, revealing how the distributions broaden and rotate in the complex plane as the interaction is approached.}
  \label{fig:field_components}
\end{figure*}

To expose the internal reorganization of complex eigenmodes we analyze the field in the \emph{quadrature space}, i.e., the Re--Im plane of the complex amplitude \(\psi(x,y)=R(x,y)+\mathrm{i}\,I(x,y)\). Although the Re and Im axes are orthogonal by construction, the sampled components \(R(x,y)\) and \(I(x,y)\) need not be statistically independent. Their weighted joint distribution therefore encodes two distinct contributions to mode delocalization: marginal broadening of each quadrature and structural reconfiguration (correlation or hybridization) between quadratures.

Figure~\ref{fig:field_components} provides the field-level decomposition associated with the spectral interaction shown in Fig.~\ref{fig:mode_evolution}.
Panels (a--c) present the real parts of $\psi(x,y)$ for {\bf Mode~1} at A--C, and (d--f) show the corresponding real parts for {\bf Mode~2} at D--F.
Panels (g--i) and (j--l) depict the imaginary components for the same parameter values.
These maps illustrate that, as the control parameter approaches the avoided crossing, the nodal structure of each mode becomes increasingly distorted and mixed, reflecting the hybridization already visible in the intensity distributions.

Panels (m--s) offer a complementary complex-plane view, where each point $(\Re\psi,\Im\psi)$ is weighted by its local intensity $|\psi|^2$.
Away from the avoided crossing the distributions cluster along nearly one-dimensional manifolds, whereas near the interaction they spread broadly across the complex plane.
This behavior makes explicit the redistribution of amplitude between real and imaginary components and signals delocalization of the eigenmodes.
Taken together, Figs.~\ref{fig:mode_evolution} and \ref{fig:field_components} reveal how spectral interactions and field decompositions consistently point to mode exchange and hybridization as hallmarks of the avoided crossing.

\section{Quadrature space analysis of complex amplitudes}
\label{sec:value_space}

We analyze the interior samples $\Omega\subset\mathbb{R}^2$ of the complex field
\begin{equation}
\psi(x,y)=R(x,y)+i\,I(x,y)
\end{equation}
with weights $w(x,y)=|\psi(x,y)|^2$. Exterior zero amplitude points are excluded. Practically, a grid point is retained if $w>0$ or if a 4-connected neighbor has $w>0$, so that thin nodal sets are not misclassified as exterior.

Because any complex eigenfunction is invariant under a global phase, $\psi'(x,y)=e^{i\chi}\psi(x,y)$, the decomposition $(R,I)$ is not unique. This transformation is equivalent to a planar rotation
\begin{equation}
\begin{pmatrix}R'\\[2pt] I'\end{pmatrix}
=\mathbf{Q}(\chi)\begin{pmatrix}R\\ I\end{pmatrix},
\qquad
\mathbf{Q}(\chi)=
\begin{pmatrix}
\cos\chi & -\sin\chi\\
\sin\chi & \ \cos\chi
\end{pmatrix},
\label{eq:global_rotation}
\end{equation}
so that comparisons across different values of the control parameter $\varepsilon$ require a consistent global-phase convention (i.e., gauge).

To establish such a convention we determine, for each $\varepsilon$, a rotation angle $\theta(\varepsilon)$ from the principal axes of the weighted covariance matrix:
\begin{equation}
\begin{aligned}
\boldsymbol{\Sigma}(\varepsilon) &=
\begin{pmatrix}
\langle R^2\rangle_w & \langle RI\rangle_w \\
\langle RI\rangle_w & \langle I^2\rangle_w
\end{pmatrix}, \\
\text{with } \langle f\rangle_w &=
\frac{\sum_{(x,y)\in\Omega} w(x,y) f(x,y)}
     {\sum_{(x,y)\in\Omega} w(x,y)} .
\end{aligned}
\end{equation}

The orientation angle is described by
\begin{equation}
\theta(\varepsilon)
=\tfrac12\,\mathrm{atan2}\!\left(2\langle RI\rangle_w,\;\langle R^2\rangle_w-\langle I^2\rangle_w\right).
\label{eq:M1_angle}
\end{equation}
This  gauge is second-moment based~\cite{Jolliffe02, GonzalezWoods08}: It aligns the \emph{quadrature space} frame with the major axis of the covariance ellipse. When the weighted mean vector $\langle(R,I)\rangle_w$ is numerically negligible, the covariance ellipse typically retains a robust orientation; hence $\theta(\varepsilon)$ provides a stable phase reference unless $\boldsymbol{\Sigma}$ is isotropic.

\begin{figure*}[htbp]
  \centering
  \includegraphics[width=17.0cm]{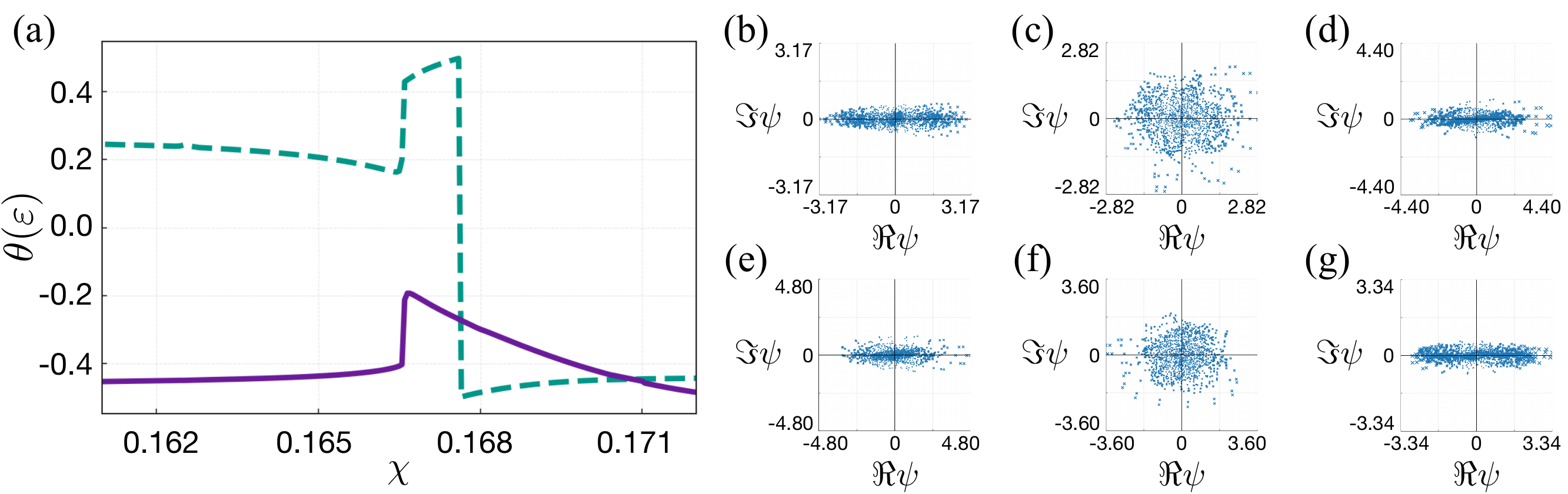}
  \caption{(a) Parameter dependence of the rotation angle $\theta(\varepsilon)$ for the two interacting modes (blue dashed: {\bf Mode~1}; purple solid: {\bf Mode~2}). Values on the vertical axis are given in units of $\pi$. The trajectories exhibit sharp variations when passing through the avoided crossing region, reflecting the rapid change in eigenmode orientation.
  (b--d) Complex-plane representations $(\Re\psi,\Im\psi)$ of {\bf Mode~1}, corresponding to Fig.~\ref{fig:field_components}(m--o).
  (e--g) Complex-plane representations of {\bf Mode~2}, corresponding to Fig.~\ref{fig:field_components}(p--r).
  These rotated distributions highlight how both modes reorganize in the complex plane, making the symmetry of the hybridization more transparent.}
  \label{fig:M1_rotation}
\end{figure*}

Figure~\ref{fig:M1_rotation}(a) plots $\theta(\varepsilon)$ for the two interacting modes (dashed: {\bf Mode~1}, solid: {\bf Mode~2}). The curves exhibit sharp variations near $\varepsilon\approx0.166$--0.167, the location of the avoided crossing, showing that the \emph{quadrature space} orientation itself undergoes a rapid reconfiguration in tandem with the spectral interaction.

In practice, we apply the $\theta(\varepsilon)$ alignment as an \emph{active} rotation of the data,
\begin{equation}
R' + iI' \;=\; e^{-\,i\,\theta(\varepsilon)}\big(R + iI\big),
\label{eq:active_rotation}
\end{equation}
so that the long axis of the covariance ellipse collapses onto the $R'$ axis. This choice makes the marginal distributions $p_{R'}$ and $p_{I'}$ directly comparable across $\varepsilon$. Panels (b--d) and (e--g) of Fig.~\ref{fig:M1_rotation} illustrate the effect: They show the complex-plane distributions $(\Re\psi,\Im\psi)$ of the two modes, rotated using the anchor value of $\theta(\varepsilon)$ depending on the control parameter $\varepsilon$. Compared to the unrotated representations in Fig.~\ref{fig:field_components}(m--r), the anchor-based rotation exposes a clearer symmetry of the hybridization, highlighting how both modes reorganize in the complex plane once aligned by a consistent gauge. This procedure ensures that subsequent information-theoretic measures are evaluated within a uniform reference frame.

The anchor-based alignment therefore provides a consistent \emph{quadrature space} gauge across $\varepsilon$, ensuring that the rapid reorientation of the modes in the avoided-crossing region is properly accounted for. With this gauge fixed, we are now in a position to analyze the statistical structure of the rotated complex amplitudes. In the next step we introduce information-theoretic measures that characterize how the \emph{quadrature space} distribution evolves across the AC, thereby quantifying the delocalization and correlations underlying the hybridization process.

\begin{figure*}[htbp]
  \centering
  \includegraphics[width=14.0cm]{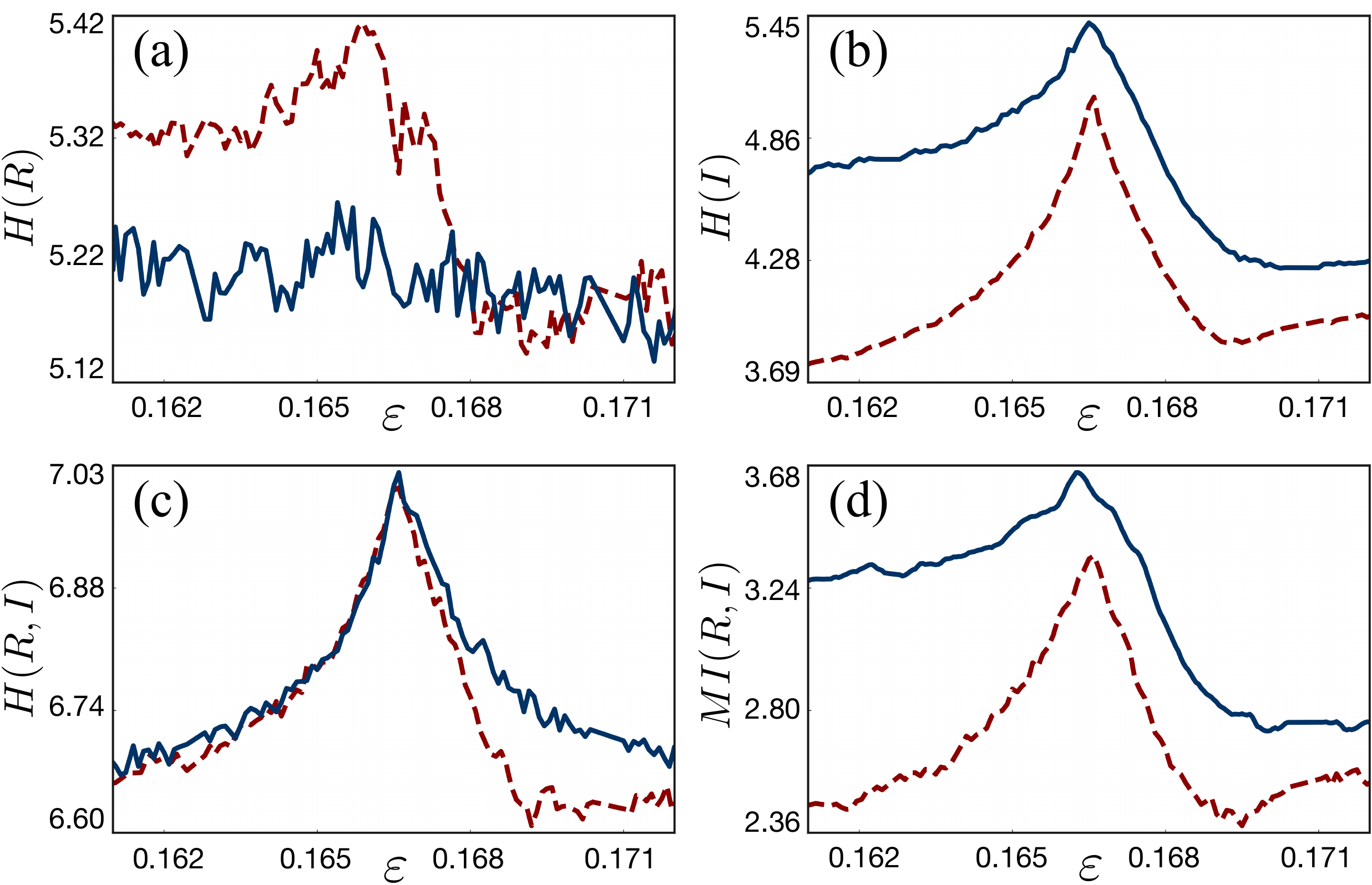}
  \caption{Information-theoretic measures of the rotated \emph{quadrature space} distributions as a function of the control parameter $\varepsilon$.
  (a) Shannon entropy of the real component $H(R)$. Although no sharp peak is observed, $H(R)$ remains consistently larger than $H(I)$ except within the avoided-crossing region.
  (b) Shannon entropy of the imaginary component $H(I)$, which exhibits a clear peak at the AC point.
  (c) Joint entropy $H(R,I)$, also showing a pronounced maximum at the AC.
  (d) Mutual information $\mathrm{MI}(R,I)$ between the two components, likewise peaking in the same region.
  Remarkably, the ratio $\mathrm{MI}(R,I)/H(R,I)$ reaches $\sim0.5$, indicating that inter-component correlations account for nearly half of the total entropy in the avoided-crossing regime.}
  \label{fig:entropy_measures}
\end{figure*}

\section{quadrature space discretization and information measures}
\label{sec:quad_space}
This section defines a compact protocol for discretizing the complex-valued field across different deformation parameters $\varepsilon$ and for comparing their information-theoretic signatures.
To compare different $\varepsilon$ on the same \emph{quadrature space} region, we anchor the global window at the avoided-crossing center $\varepsilon_\star\simeq 0.16655$. From a representative AC dataset, we evaluate the anchor angle
\begin{align}
\theta_{\mathrm{anchor}} := \theta(\varepsilon_\star),
\end{align}
and rotate a sparse representative subset by this angle to estimate robust global limits
\begin{align}
(R_{\min},R_{\max})\times(I_{\min},I_{\max}),
\end{align}
using central quantiles with a light padding. This AC-anchored global window ensures reproducible, cross-dataset comparisons in a common frame. Throughout what follows, all statistics are computed in this frame.

We discretize the window into NB$\times$NB bins (NB $=500$ in practice) by defining the bin indices as
\begin{align}
a &= \Big\lfloor \tfrac{R'-R_{\min}}{\Delta R}\,\mathrm{NB}\Big\rfloor,\quad\mathrm{and}\\
b &= \Big\lfloor \tfrac{I'-I_{\min}}{\Delta I}\,\mathrm{NB}\Big\rfloor
\end{align}
with clamping $0\le a,b\le \mathrm{NB}-1$, where $\Delta R:=R_{\max}-R_{\min}$ and $\Delta I:=I_{\max}-I_{\min}$. Here $\lfloor \cdot \rfloor$ denotes the floor function (Gauss’s notation for the integer part), which always rounds down to the nearest integer. Clamping guarantees array safety for boundary samples (e.g.\ $R'=R_{\max}$) and also protects against small floating-point excursions outside the nominal interval. We adopt the standard half-open bin convention so that adjacent bins do not overlap, and the final bin is made to include the upper endpoint by the clamping rule.

Given the rotated cloud $\{(R'_j,I'_j),w_j\}$, the weighted bin counts are
\begin{align}
h_{a,b}=\sum_{j\in\Omega} w_j\,\chi_{\text{bin}(a,b)}\!\big(R'_j,I'_j\big),
\end{align}
where $\chi_{\text{bin}(a,b)}$ is the characteristic function of bin $(a,b)$, equal to one if $(R'_j,I'_j)$ lies inside that bin and zero otherwise. The total weight and the normalized probabilities are stated compactly as
$$
Z=\sum_{a,b} h_{a,b},\qquad p_{a,b}=\frac{h_{a,b}}{Z},\qquad \sum_{a,b} p_{a,b}=1.
$$
From these, the marginals follow as $p_R(a)= \sum_b p_{a,b}$ and $p_I(b)= \sum_a p_{a,b}$.

Since all distributions are expressed in the rotated coordinates, we hereafter omit primes on $(R',I')$ for clarity. Using natural logarithms (nats), the entropies are defined as
\begin{align}
H_R &= -\sum_{a} p_R(a)\ln p_R(a),\\
H_I &= -\sum_{b} p_I(b)\ln p_I(b),\\
H_{R,I} &= -\sum_{a,b} p_{a,b}\ln p_{a,b},
\end{align}
and the \emph{mutual} information is given by
\begin{align}
\mathrm{MI}(R,I) = H_R+H_I-H_{R,I}.
\label{eq:MI}
\end{align}

In this framework, Shannon entropy quantifies the delocalization of the \emph{quadrature space} distributions, whereas the mutual information captures the statistical dependence between the two quadratures. Figure~\ref{fig:entropy_measures} presents the resulting diagnostics. Panel (a) shows $H(R)$, which does not develop a sharp maximum; outside the avoided crossing $H(R)$ consistently exceeds $H(I)$, but its contribution to the peak structure is limited. Panel (b) shows $H(I)$, which develops a pronounced peak at $\varepsilon\simeq\varepsilon_\star$, while panel (c) shows that $H(R,I)$ likewise peaks at the same location. Panel (d) shows that $\mathrm{MI}(R,I)$ is also maximized at $\varepsilon_\star$, indicating that the two quadratures are most strongly correlated in the hybridization region. Importantly, the peak of $H(R,I)$ is predominantly driven by the enhancement of $H(I)$ together with the surge of $\mathrm{MI}(R,I)$, rather than by $H(R)$, which remains comparatively structureless.

This conclusion is geometrically consistent with the rotated clouds in the complex plane. Far from the avoided crossing the distributions are elongated mainly along the $R$ axis, so that knowledge of $R$ conveys little information about $I$ and the mutual information is small. Near $\varepsilon_\star$, however, the cloud becomes nearly circular, indicating comparable fluctuations in $R$ and $I$ and a concomitant rise in their correlation. Quantitatively, the ratio $\mathrm{MI}(R,I)/H(R,I)$ reaches values of order $0.5$ for both mode families, i.e., nearly half of the joint entropy in the avoided-crossing regime is accounted for by inter-component correlations. Thus the entropy growth across the avoided crossing is not merely a delocalization effect but is dominantly correlation driven a mechanism that would remain hidden if one only analyzed the conventional Shannon entropy of the intensity $\rho(x,y)=|\psi(x,y)|^{2}$ and its spatial marginals.

Unless otherwise stated, weights are taken to be the intensities $w=|\psi|^2$ both in the covariance that defines $\theta(\varepsilon)$ and in the \emph{quadrature space} histogram $h_{a,b}$. For comparison we also computed an unweighted variant using unit weights. In practice, the weighted choice sharpens the peak structures in $H_I$ and $\mathrm{MI}$ near the avoided crossing, whereas the unweighted case produces the same qualitative behaviour but with less pronounced maxima. We further tested robustness against the number of bins (NB $=300,500,700$): the overall trends and peak positions remain unchanged, with only a trivial rescaling of absolute values relative to the NB $=500$ reference. Together with the AC-anchored global frame, these checks confirm that our conclusions do not hinge on technical choices of weighting or discretization.

\section{Conclusion}
\label{sec:conclusion}
We have presented an information-theoretic framework operating in the \emph{quadrature space} of complex eigenmodes, which complements conventional intensity-based analyses. Instead of using only the spatial intensity \( |\psi(x,y)|^2 \), our method assigns probabilistic weights to the real and imaginary quadratures and examines their joint statistics, revealing internal reorganizations that intensity alone conceals.

When applied across a control parameter sweep, the quadrature space distributions show that avoided crossings produce delocalization resulting from both marginal broadening and structural redistribution between quadratures. The procedure is phase-consistent, robust to nodal sets, requires no external phase reference, and extends naturally to other non-Hermitian regimes and to neighborhoods of exceptional points.

\section{Acknowledgement}
This work was supported by the National Research Foundation of Korea (NRF) through a grant funded by the Ministry of Science and ICT (Grants Nos. RS-2023-00211817 and RS-2025-00515537), the Institute for Information \& Communications Technology Promotion (IITP) grant funded by the Korean government (MSIP) (Grants Nos. RS-2019-II190003 and RS-2025-02304540), the National Research Council of Science \& Technology (NST) (Grant No. GTL25011-401), and the Korea Institute of Science and Technology Information (Grant No. P25026). S.L. acknowledges support from the National Research Foundation of Korea (NRF) grants funded by the MSIT (Grant No. RS-2024-00432214).

\section*{Data availability}
Data underlying the results presented in this paper are not publicly available but may be obtained from the authors upon reasonable request.

\section*{Declaration of competing interest}
The authors declare that they have no known competing financial interests or personal relationships that could have appeared to influence the work reported in this paper.


\end{document}